\documentclass[runningheads]{llncs}
\usepackage{graphicx}
\usepackage{customization}
\usepackage[linesnumbered,ruled,vlined]{algorithm2e}
\usepackage{mathtools}
\usepackage{algpseudocode, multirow, booktabs}
\usepackage[breaklinks,colorlinks]{hyperref}

\hypersetup{
    linkcolor=eq-fig-tab-color,  
    citecolor=ref-color,  
}
\begin{document}
\title{Poisson2Sparse: Self-Supervised Poisson Denoising From a Single Image}
%
%
\author{Calvin-Khang Ta$^{1,}$\thanks{joint first authors},
Abhishek Aich$^{1,\star}$,
Akash Gupta$^{2,\star}$,
Amit K. Roy-Chowdhury$^1$}
\authorrunning{C.K. Ta et al.}
%
\institute{$^1$University of California, Riverside, $^2$Vimaan AI\\
\email{\{cta003@, aaich001@, agupt013@, amitrc@ece.\}ucr.edu}
}

\maketitle              
\begin{abstract}
Image enhancement approaches often assume that the noise is signal independent, and approximate the degradation model as zero-mean additive Gaussian. However, this assumption does not hold for biomedical imaging systems where sensor-based sources of noise are proportional to signal strengths, and the noise is better represented as a Poisson process. In this work, we explore a sparsity and dictionary learning-based approach and present a novel self-supervised learning method for single-image denoising where the noise is approximated as a Poisson process, requiring no clean ground-truth data. Specifically, we approximate traditional iterative optimization algorithms for image denoising with a recurrent neural network that enforces sparsity with respect to the weights of the network. Since the sparse representations are based on the underlying image, it is able to suppress the spurious components (noise) in the image patches, thereby introducing implicit regularization for denoising tasks through the network structure. Experiments on two bio-imaging datasets demonstrate that our method outperforms the state-of-the-art approaches in terms of PSNR and SSIM. Our qualitative results demonstrate that, in addition to higher performance on standard quantitative metrics, we are able to recover much more subtle details than other compared approaches. Our code is made publicly available at \url{https://github.com/tacalvin/Poisson2Sparse}
\end{abstract}


\section{Introduction}

Biomedical image denoising is a challenging inverse problem of recovering a clean noise-free image from its corresponding corrupted version. Current denoising strategies \cite{zamir2021multi, Chen_2021_CVPR} assume that most prevalent noisy images can be modeled with additive Gaussian noise. Though this assumption shows reasonable performance for some applications, it is physically unrealistic for biomedical images as the noise varies proportionally to the signal strength and is signal-dependent \cite{hasinoff2014photon}. As the image acquisition process is discrete in nature, the data captured by the imaging sensors are often corrupted by shot noise which can be modeled as a Poisson Process \cite{Pawley2006, doi:10.1152/physiol.00036.2006}. Thus, Poisson denoising methods are of utmost importance in biomedical image processing and analysis. 

Supervised deep learning approaches are proven to be effective for image denoising tasks \cite{zamir2021multi, Chen_2021_CVPR} and are mostly developed on sophisticated neural networks. The performance of such approaches heavily relies on the availability of large datasets to train the networks. These large-scale datasets often involve data pairs of a clean image, $\bm{X}$, and a noisy image, $\noisyxmat$. 
However in most practical settings, such as bio-imaging \cite{aldaz2010live}, it is difficult to obtain such data pairs. Hence in order to tackle such scenarios, self-supervised \textit{internal learning} methods \cite{8579082} have been introduced that attempt to employ randomly-initialized neural networks to capture low-level image statistics. Such networks are used as priors to solve standard inverse problems such as denoising for \textit{single} images without requiring its clean counterpart.\\[1em]
%
\textbf{Related Works.} The general supervised approach for using deep learning based denoisers is to use Convolutional Neural Networks (CNNs) and given a dataset of clean and noisy images, to learn the mapping between them \cite{zamir2021multi, Aich_2020_CVPR, Chen_2021_CVPR, gupta2020deep}. Recent works in self-supervised learning \cite{lehtinen2018noise2noise, 8954066, xu2021deformed2self} have shown that even without the use of explicit ground truth, deep learning models can offer surprisingly strong performance in a range of tasks. \calvin{Our work is largely inspired by works such as the Deep Image Prior \cite{8579082}  or Self2Self \cite{quan2020self2self}  in which a deep learning model is trained directly on a single test image with no additional datasets. The core assumption of such works is that the network is implicitly acting as the regularizer and empirical results show the network resists fitting to noise.}
Given such advantages of internal learning methods, \emph{we propose a novel self-supervised approach to denoise biomedical images that follow the Poisson noise model.} Building on the implicit regularization of a network, we utilize sparse representations through the network design to handle Poisson Noise. Different from prior works \cite{zamir2021multi, Chen_2021_CVPR}, \emph{we consider the scenario where only a single noisy image is collected with \underline{no} corresponding ground-truth clean image \calvin{and the noise is Poisson distributed.}} We employ the sparse representations because, intuitively, an image will contain many recurrent patches and by finding a sparse representation of the image patches, we can represent it as a linear combination of elements in a dictionary. By utilizing this sparse coding framework, the dictionary elements will \calvin{contain the basis vectors} which minimize the reconstruction error and suppress the noise. \\[1em]
\textbf{Contributions.} 
We present a novel method for single image Poisson denoising which leverages a neural network's inductive bias. Specifically, we use the neural network to learn the dictionary elements that are resistant to the input noise by modeling similar and repetitive image patches. We then utilize sparse representations to reconstruct the image through learned dictionaries and further suppress input noise. By leveraging the internal learning strategy, our method gains two key advantages: \textit{1)} we only need a single noisy input image which is desirable in many biomedical applications where data is scarce, and \textit{2)} we can train our model in an entirely self-supervised manner with no ground-truth which makes our approach extremely practical. 
Experiments show our approach is able to outperform existing state-of-the-art methods significantly. We illustrate our method in Figure \ref{fig:main_fig} and summarize the same in Algorithm 1 in the supplemental material.

\section{Proposed Methodology: Poisson2Sparse}
\textbf{Problem Statement.} Given a single Poisson noisy input image, we aim to generate the corresponding clean image. We propose to utilize the patch recurrence property in an image to learn a dictionary that can be utilized to generate a sparse representation such that the clean image can be recovered. The dictionary elements will ideally represent a set of over-complete basis functions that can well represent the image. The sparse representation is then used to reconstruct the image using the learned dictionary to suppress the noise present in the input image \cite{salmon2014poisson, 6918528}.
Let the Poisson noisy image be represented by  $\noisyxmat \in \mathcal{R}^{d \times d}$ and its corresponding vectorized form denoted by $\noisyx\in \mathcal{R}^{d^2}$, where $d$ is the image dimension size.  \calvin{Our objective is to learn a dictionary to obtain sparse representation of $\bm{\noisyxmat}$ that can be utilized to recover an estimate of the clean image $\widehat{\bm{X}}$}. To this end, we incorporate an unrolled iterative optimization algorithm approximated using a neural network to learn the dictionary and decompose the noisy image representation into a sparse code. \calvin{In this section, we first derive the dictionary-based sparse representation learning algorithm for the vectorized image}, where we represent the learnable dictionary as $\mathcal{D} \in \mathcal{R}^{d^2 \times k}$ and the sparse vector as $\bm{\alpha} \in \mathcal{R}^{k \times 1}$. Here, $k$ is the number of elements in $\bm{\alpha}$. Next, we leverage the Convolution Sparse Coding model~\cite{6618901} to obtain an optimization solution where the dictionary is independent of the vectorized image dimension. This is done by changing the application of the dictionary $\mathcal{D}\bm{\alpha}$ from a matrix-vector product into a convolution with $M$ dictionaries $D\in \mathcal{R}^{k \times k}$ around $M$ sparse feature maps $\mathrm{A} \in \mathcal{R}^{d \times d}$ \cite{6618901}. \\[1em]
\textbf{Poisson-based Optimization Regularizer.} We assume that the pixel values of the noisy vectorized image $\noisyx$ are Poisson distributed random variables parameterized by pixel values of \calvin{the ground truth vectorized image} $\bm{x}$ at every $i^{th}$ index of the image. This allows $\noisyx[i]$ to be modeled as $\noisyx[i] \sim \mathcal{P}(\bm{x}[i])$ \cite{salmon2014poisson, 6918528} where $\mathcal{P}$ is the Poisson process defined as follows:
\begin{align}
    \mathcal{P}_{\bm{x}[i]}(\noisyx[i]) = \dfrac{\bm{x}[i]^{\noisyx[i]} \exp(-\bm{x}[i])}{\noisyx[i]!}
    \label{eq:pdist}
\end{align}
In order to estimate a denoised clean vector $\estimatex$, we maximize the log-likelihood of \eqref{eq:pdist}. The maximum log-likelihood estimation for clean vector recovery is performed by minimizing the following optimization problem \cite{5462884}
\calvin{
\begin{align}
    \min_{\bm{x}} \big{(}\mathbbm{1}^T\bm{x} - \noisyx^T \log(\bm{x})\big{)} \text{ s.t. } \bm{x} \succcurlyeq \bm{0}
    \label{eq:log-like-pdis}
\end{align}
}where $\mathbbm{1}\in \mathcal{R}^{d^2}$ is a vector of ones, $\succcurlyeq$ denotes element-wise inequality, and $\log(\cdot)$ is applied element wise. However, the optimization problem defined in \eqref{eq:log-like-pdis} is known to be an ill-posed problem \cite{hansen1998rank, tikhonov1995numerical}. In order to address this, we follow a sparse representation approach \cite{5459452} and aim to estimate $\estimatex$ by computing an $s$-sparse vector $\alpha$ and a dictionary $\mathcal{D}$ such that $\estimatex=\mathcal{D}\alpha$.
\begin{figure}[t]
    \centering
    \includegraphics[width=0.95\textwidth]{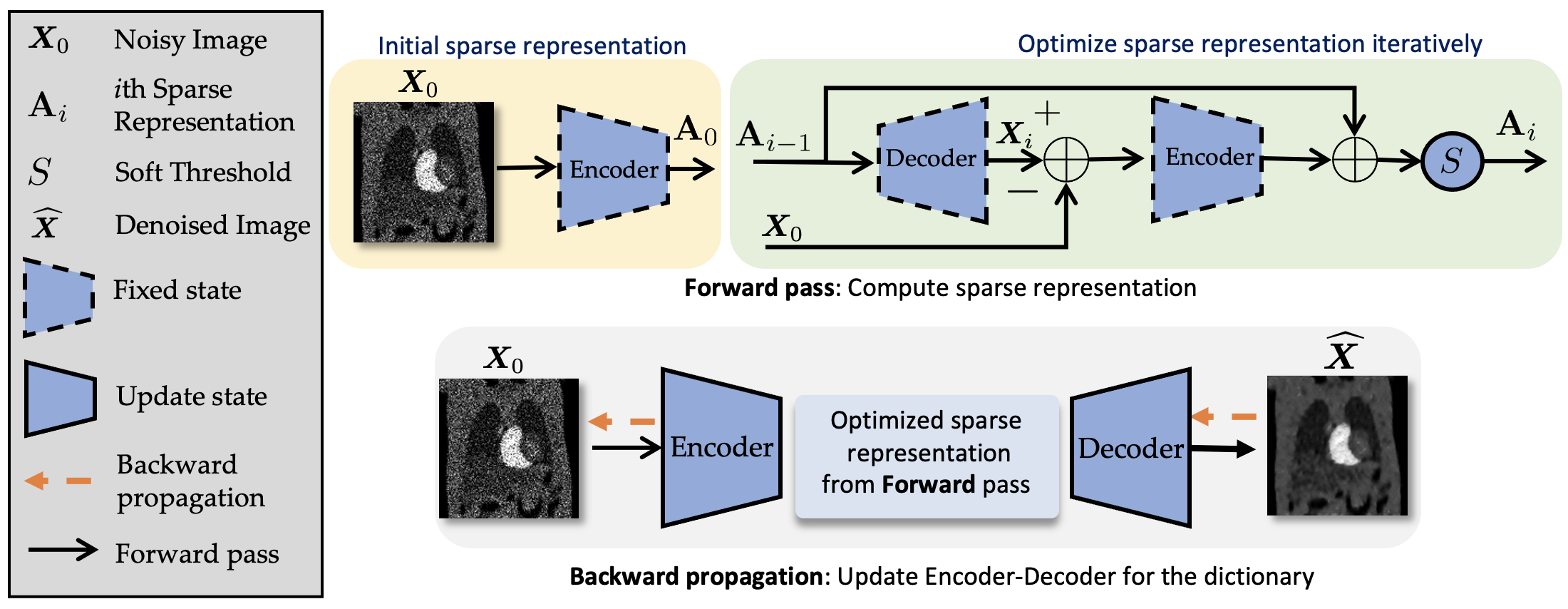}
    \caption{\textbf{Method Overview.} Poisson2Sparse Optimization Steps: We optimize the sparse representation in the forward pass and then we update the encoder and decoder through back-propagation in an alternating manner.}
    \label{fig:main_fig}
\end{figure}
\begin{align}
    \begin{aligned}
        \min_{\bm{\mathcal{D},\alpha}} \big{(}\mathbbm{1}^T(\mathcal{D}\bm{\alpha}) - \noisyx^T \log(\mathcal{D}\bm{\alpha})\big{)}~~\text{s.t.}~ ||\bm{\alpha}||_0 \leq s, \quad \mathcal{D}\bm{\alpha} \succcurlyeq \bm{0}
        \label{eq:col-opt-prob-2}
    \end{aligned}
\end{align}
This optimization problem in \eqref{eq:col-opt-prob-2} is further relaxed by setting $\mathcal{D}\bm{\alpha} = \exp(\mathcal{D}\bm{\alpha})$ to handle the non-negativity constraint \cite{salmon2014poisson}.
Furthermore, the $\ell_0$ constraint on $\bm{\alpha}$ makes \eqref{eq:col-opt-prob-2} an NP-hard problem \cite{ge2011note}. Hence, we use the $\ell_1$ relaxation as in \cite{6020799}, resulting in the following problem:
\begin{align}
    \begin{aligned}
        \min_{\mathcal{D},\bm{\alpha}} \big{(}\mathbbm{1}^T\exp(\mathcal{D}\bm{\alpha}) - \noisyx^T (\mathcal{D}\bm{\alpha})\big{)} + \lambda ||\bm{\alpha}||_1
    \end{aligned}
    \label{eq:ista_min_problem}
\end{align}

The above optimization problem in \eqref{eq:ista_min_problem} now estimates $\estimatex$ by solving for sparse prior $\bm{\alpha}$ and dictionary $\mathcal{D}$. Inspired from the Iterative Shrinkage Thresholding Algorithm (ISTA) \cite{daubechies2004iterative}, we propose to solve for $\bm{\alpha}$ and $\mathcal{D}$ in an alternating 
manner using a neural network based approach. In order to solve for $\alpha$, traditional approaches \cite{daubechies2004iterative} solve the following optimization problem $$\min |\bm{\alpha}|_0 \quad\text{s.t}\quad \bm{x} = \mathcal{D}\bm{\alpha}$$
However, trying to solve for that objective directly is difficult and a common approximation is the following objective using the $\ell_1$ relaxation.
\begin{align}
    \text{arg}\min_{\bm{\alpha}} \bigg{(}\dfrac{1}{2} ||\bm{x}-\mathcal{D}\bm{\alpha}||_{2}^{2} + \lambda|\bm{\alpha}|_1\bigg{)}
    \label{eq:ista_objective}
\end{align}
The ISTA algorithm aims to solve \eqref{eq:ista_objective} via the update step $\bm{\alpha} \leftarrow S(\bm{\alpha} + \frac{1}{L}\mathcal{D}^T(\noisyx- \mathcal{D}\bm{\alpha}))$, where $L \leq \sigma_\text{\text{max}}(\mathcal{D}^T\mathcal{D})$ and $S$ is the soft threshold operator with a threshold of $\epsilon$ defined as $S_\epsilon(\bm{x}) = \text{sign}(\bm{x})\text{max}(|\bm{x}| - \epsilon,0)$.
With this update step, the ISTA algorithm iteratively refines the computed sparse code until a specified convergence criterion. However, \calvin{in our problem formulation we work with images with dimension $d$} \calvin{making the dictionary $\mathcal{D} \in \mathcal{R}^{d^{2}\times k}$ dependent} on the size of the input image. To address this, we use the Convolutional Sparse Coding model as in \cite{simon2019rethinking} and replace the matrix-vector product with a convolution \calvin{(denoted by $*$)}:  
\begin{align}
     \mathcal{D}\bm{\alpha} = \sum_{j}^{M} D_j * \mathrm{A}_j = \bm{D} * \bm{\mathrm{A}}
\end{align}
where $D_j\in\mathcal{R}^{k\times k}$ is filter convolved around a sparse feature map $\mathrm{A}_j\in \mathcal{R}^{d\times d}$. This new form of the sparse code and application of the dictionary decouples the size of the dictionary from the input image size and removes the need to scale the model with respect to the image size and gives us the following update step:
\begin{align}
    \bm{\mathrm{A_i}} \leftarrow S\big{(}\bm{\mathrm{A_{i-1}}} + \bm{D}^T * (\bm{\noisyxmat}- \bm{D} * \bm{\mathrm{A_{i-1}}})\big{)}
    \label{eq:csc_ista}
\end{align}





Using the ISTA approach, we can remove the need to optimize for $\bm{\alpha}$ in \eqref{eq:ista_min_problem} and rewrite the objective function, where $\odot$ is the Hadamard product, as
\begin{align}
    \begin{aligned}
        \min_{D} \big{(}\exp( \bm{D} *\bm{\mathrm{A}}) - \noisyxmat \odot (\bm{D} * \bm{\mathrm{A}})\big{)}
    \end{aligned}
    \label{eq:csc_ista_min_problem}
\end{align}
\textbf{Poisson2Sparse Algorithm.} 
To solve \eqref{eq:csc_ista_min_problem} using the ISTA algorithm, we represent $\bm{D} * \bm{\mathrm{A}}$ using a neural network $\bm{f_\theta}$. The network $\bm{f_\theta}$ which contains a single encoder and decoder which computes the sparse representation $\bm{\mathrm{A}}$ with respect to the network parameters which allows for learnable dictionaries $\bm{D}$ through back-propagation. Mathematically, this can be represented as $\bm{D} * \bm{\mathrm{A}}= \bm{f_\theta}(\noisyxmat)$.
The modified optimization problem \eqref{eq:csc_ista_min_problem}, where we aim to enforce sparsity through the network structure implicitly, can be rewritten as 
\begin{align}
    \begin{aligned}
        \min_{\bm{\theta}} \big{(}\exp(\bm{f_\theta}(\noisyxmat)) - \noisyxmat \odot \bm{f_\theta}(\noisyxmat)\big{)} 
    \end{aligned}
    \label{eq:nn_ista_min_problem}
\end{align}
which we will refer to as $\mathcal{L}_{Poisson}$ moving forward. Inspired by recent works in internal learning \cite{8579082}, we aim to use network $\bm{f_\theta}$ 
that will implicitly enforce sparsity through the network's structure. Hence, \emph{we propose to adapt the internal learning approach \textit{via} the ISTA algorithm}. Prior works \cite{simon2019rethinking} have demonstrated that by  we can approximate computing the update step in the ISTA algorithm by replacing the $D$ and term $D^T$ with a decoder and encoder respectively in \eqref{eq:csc_ista}.
\begin{align}
    \bm{\mathrm{A}} \leftarrow S\big{(}\bm{\mathrm{A}}+ Encoder(\noisyxmat- Decoder(\bm{\mathrm{A}}))\big{)}
    \label{eq:cscnet_iteration}
\end{align}
Following \cite{simon2019rethinking}, we approximate ISTA to a finite $T$ number of iterations and by approximating the traditional update with \eqref{eq:cscnet_iteration} which naturally lends itself to a recurrent neural network-like structure where instead of passing in a $T$ length sequence we are refining the sparse code $\bm{\mathrm{A}}$ over $T$ steps instead. This results in the network's forward pass that computes the sparse code as well as the application of the dictionary to the computed sparse code. The sparse code is then refined over a finite number of steps as opposed to being run until convergence in the traditional ISTA algorithm.\\[1em]
\textbf{Poisson2Sparse Training.} In order to train the network, we follow \cite{Huang_2021_CVPR} and generate our input and target image pairs by using the random neighbor down-sampling. The random neighbor down-sampling is done by dividing the input image into $k\times k$ blocks and for each block, two adjacent pixels are selected and are used to create down-sampling functions denoted by $g_1$ and $g_2$. This down-sampling approach avoids learning the trivial identity solution where the network learns to simply map the input to output to solve \eqref{eq:nn_ista_min_problem}. It creates image pairs that are similar in appearance but are slightly different in terms of the ground truth pixels. These image pairs can be seen as different realizations of the same noise distribution and by trying to minimize the loss function for these noisy pairs we can estimate the true ground truth signal. Optimizing \eqref{eq:nn_ista_min_problem} alone with the generated image pairs will result in blurry images due to the slight difference in the underlying ground truth and thus we incorporate the regularizer proposed in \cite{Huang_2021_CVPR} which is defined as the following with strength $\bm{\mu_N}$.
 \begin{align}
     \mathcal{L}_{\text{N}} = \Vert \bm{f_\theta}(g_{1}(\noisyxmat)) - g_2(\noisyxmat) - g_1(\bm{f_\theta}(\noisyxmat)) - g_2(\bm{f_\theta}(\noisyxmat))\Vert^2_2
\end{align}
We also add an L1 reconstruction loss which is referred as $\mathcal{L}_{L1}$ and it aids in the reduction of artifacts generated from \eqref{eq:nn_ista_min_problem}. The overall loss function used to optimize the dictionary parameters $\bm{f_\theta}$ which is applied per pixel is given as: 
\begin{align}
    \mathcal{L} = \mathcal{L}_{Poisson} + \mathcal{L}_{L1} + \bm{\mu_{\text{N}}}\mathcal{L}_{\text{N}}
    \label{eq:full_loss}
\end{align}


\section{Experiments and Results}
\begin{table}[t]
\centering
\begin{minipage}{0.55\textwidth}
\renewcommand{\arraystretch}{1.2}
\resizebox{0.995\textwidth}{!}{
\begin{tabular}{c|c|c|c|c|c}
\hline
\multirow{2}{*}{\textbf{Dataset}} & \multirow{2}{*}{$\lambda$} & \textbf{DIP} & \textbf{Self2Self} & \textbf{BM3D} & \textbf{Ours} \\ \cmidrule(l){3-6} 
        &    & \multicolumn{4}{c}{\textbf{PSNR (dB)/SSIM}}                    \\ 
\hline
\multirow{3}{*}{PINCAT}           & 40                         & 30.222/0.878 & 33.138/0.942       & 32.553/0.944  & \textbf{34.309}/\textbf{0.957}  \\
        & 20 & 26.495/0.765 & 30.067/0.893 & 31.448/0.911 & \textbf{32.202}/\textbf{0.937} \\
        & 10 & 22.495/0.601 & 27.028/0.814 & 27.961/0.867 & \textbf{30.005}/\textbf{0.898} \\ 
\hline
FMD     & -  & 32.417/\textbf{0.916} & 28.563/0.688 & 29.627/0.854 & \textbf{32.980}/0.897 \\ 
\hline
\rowcolor{yellow!10}
Average & -  & 27.907/0.790 & 29.699/0.834 & 30.397/0.894 & \textbf{32.374}/\textbf{0.922}\\ 
\hline
\end{tabular}}
\end{minipage}
\begin{minipage}{0.4\textwidth}
\caption{\textbf{Quantitative Results} show that we outperform both self-supervised works (DIP, Self2Self) and traditional work BM3D (best results in bold).}
\label{table:experiment_results}
\end{minipage}

\end{table}
\textbf{Datasets and Experimental Setup.} We used the test set of the following datasets to evaluate our approach. \textit{(a)} \textbf{Florescent Microscopy Denoising Dataset} (FMD) \cite{zhang2019poisson}: This dataset contains 12000 Fluorescence Microscopy images of size $512 \times 512$ containing noise where Poisson noise is dominant. The images contain real noise of varying strengths from across multiple microscopy modalities. In order to obtain clean images a common technique is to average multiple images for an effective ground truth. For this dataset this is especially appropriate as the samples are static which allows for accurate registration between the 50 images used to obtain the effective ground truth. \textit{(b)} \textbf{PINCAT Dataset} \cite{sharif2007adaptive}: This is a simulated cardiac perfusion MRI dataset with clean images of size $128\times 128$. For this dataset, we added artificial noise (where the strength $\lambda$  of noise indicates the maximum event count, with a lower $\lambda$ indicating a higher noise). Specifically we can represent a noisy image as $\noisyxmat = \frac{Z}{\lambda}$ where $Z\sim\mathcal{P}(\lambda X)$ and we tested with $\lambda=[40,20,10]$ following prior works \cite{xu2021deformed2self}.\\[1em]
\textbf{Baselines and Evaluation Metrics.} We use a traditional denoising approach, BM3D \cite{2006SPIE.6064..354D}, and internal learning based approaches such as Deep Image Prior (DIP) \cite{8579082} and Self2Self \cite{quan2020self2self} as our baselines. While comparison to vanilla ISTA is also appropriate we found that the results were far worse than any of the methods previously mentioned despite its fast runtime and will leave those experiments in the supplementary material. In order to compare methods we use Peak Signal to Noise Ratio (PSNR) and the Structural Similarity Index (SSIM) as evaluation metrics.\\[1em]
\textbf{Training details.} To optimize the network, we use the Adam optimizer \cite{Kingma2015AdamAM} with a learning rate of $0.0001$ and set $\bf{\mu}_N=2$ as done in \cite{Huang_2021_CVPR}. We use a single convolutional layer for both the encoder and decoder with a kernel size of $3 \times 3$, stride of $1$, and each layer has $512$ filters. For the number of steps in the network, we empirically find that $10$ gave us the best performance and we train the network over 5500 iterations. We used PyTorch for our implementation and the model was trained using a Nvidia RTX 2080 Ti for a total of 7 minutes for an image of size $128\times 128$. \\[1em]
\begin{figure*}[t!]
\begin{subfigure}[b]{0.16\textwidth} 
    \centering
    \includegraphics[width=\linewidth]{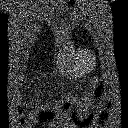}\\
    16.920 dB
    \includegraphics[width=\linewidth]{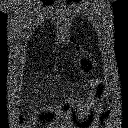}\\
    17.580 dB
    \includegraphics[width=\linewidth]{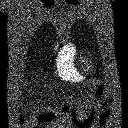}
    17.050 dB
    \begin{center}
    \textcolor{black}{(a) Noisy}
    \end{center}
\end{subfigure}%
\hfill
\begin{subfigure}[b]{0.16\textwidth} 
    \centering
    \includegraphics[width=\linewidth]{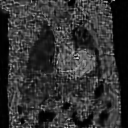}\\
    21.985 dB
    \includegraphics[width=\linewidth]{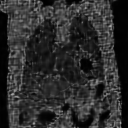}\\
    23.386 dB
    \includegraphics[width=\linewidth]{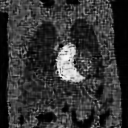}
    22.140 dB
    \begin{center}
    \textcolor{black}{(b) DIP}
    \end{center}
\end{subfigure}%
\hfill
\begin{subfigure}[b]{0.16\textwidth} 
    \centering
    \includegraphics[width=\linewidth]{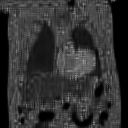}\\
    26.827 dB
    \includegraphics[width=\linewidth]{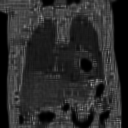}
    27.700 dB
    \includegraphics[width=\linewidth]{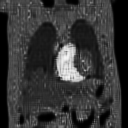}
    26.448 dB
    \begin{center}
    \textcolor{black}{(c) Self2Self}
    \end{center}
\end{subfigure}%
\hfill
\begin{subfigure}[b]{0.16\textwidth} 
    \centering
    \includegraphics[width=\linewidth]{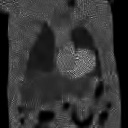}\\
    28.831 dB
    \includegraphics[width=\linewidth]{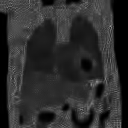}\\
    29.082 dB
    \includegraphics[width=\linewidth]{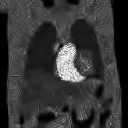}
    26.698 dB
    \begin{center}
    \textcolor{black}{(d) BM3D}
    \end{center}
\end{subfigure}%
\hfill
\begin{subfigure}[b]{0.16\textwidth} 
    \centering
    \includegraphics[width=\linewidth]{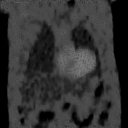}\\
    \bf{30.433 dB}
    \includegraphics[width=\linewidth]{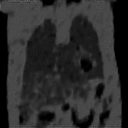}\\
    \bf{31.494 dB}
    \includegraphics[width=\linewidth]{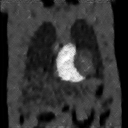}
    \bf{28.882 dB}
    \begin{center}
    \textcolor{black}{(e) Ours}
    \end{center}
\end{subfigure}%
\hfill
\begin{subfigure}[b]{0.16\textwidth} 
    \centering
    \includegraphics[width=\linewidth]{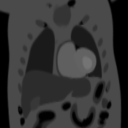} \\
    PSNR
    \includegraphics[width=\linewidth]{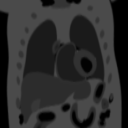} \\
    PSNR
    \includegraphics[width=\linewidth]{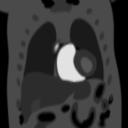} \\
    PSNR
    \begin{center}
    \textcolor{black}{(f) GT}
    \end{center}
\end{subfigure}%
    \caption{\textbf{Denoising results on the PINCAT dataset}. Noise is $\lambda=10$. We can see that our method is able to recover finer details such as the ribs in the image.}
\label{fig:PINCAT}
\end{figure*} 
\begin{figure*}[t!]
\begin{subfigure}[b]{0.16\textwidth} 
    \centering
    \includegraphics[width=\linewidth]{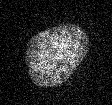} \\
    24.701 dB
    \includegraphics[width=\linewidth]{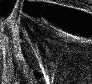} \\
    28.124 dB
    \includegraphics[width=\linewidth]{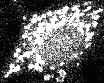} \\
    33.271 dB
    \begin{center}
    \textcolor{black}{(a) Noisy}
    \end{center}
\end{subfigure}%
\hfill
\begin{subfigure}[b]{0.16\textwidth} 
    \centering
    \includegraphics[width=\linewidth]{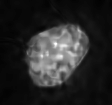} \\
    32.430 dB
    \includegraphics[width=\linewidth]{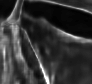} \\
    30.524 dB
    \includegraphics[width=\linewidth]{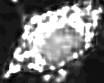} \\
    34.037 dB
    \begin{center}
    \textcolor{black}{(b) DIP}
    \end{center}
\end{subfigure}%
\hfill
\begin{subfigure}[b]{0.16\textwidth} 
    \centering
    \includegraphics[width=\linewidth]{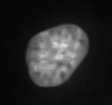} \\
    32.888 dB
    \includegraphics[width=\linewidth]{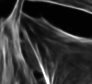} \\
    31.591 dB
    \includegraphics[width=\linewidth]{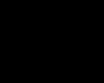} \\
    23.984 dB
    \begin{center}
    \textcolor{black}{(c) Self2Self}
    \end{center}
\end{subfigure}%
\hfill
\begin{subfigure}[b]{0.16\textwidth} 
    \centering
    \includegraphics[width=\linewidth]{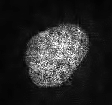} \\
    27.752 dB
    \includegraphics[width=\linewidth]{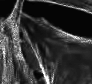}\\
    30.417 dB
    \includegraphics[width=\linewidth]{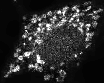}\\
    34.158 dB
    \begin{center}
    \textcolor{black}{(d) BM3D}
    \end{center}
\end{subfigure}%
\hfill
\begin{subfigure}[b]{0.16\textwidth} 
    \centering
    \includegraphics[width=\linewidth]{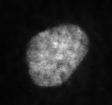} \\
    \bf{33.731 dB}
    \includegraphics[width=\linewidth]{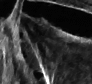} \\
    \bf{31.706 dB}
    \includegraphics[width=\linewidth]{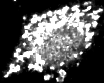} \\
    \bf{38.707 dB}
    \begin{center}
    \textcolor{black}{(e) \textbf{Ours}}
    \end{center}
\end{subfigure}%
\hfill
\begin{subfigure}[b]{0.16\textwidth} 
    \centering
    \includegraphics[width=\linewidth]{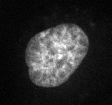} \\
    PSNR
    \includegraphics[width=\linewidth]{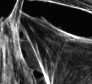} \\
    PSNR
    \includegraphics[width=\linewidth]{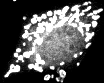} \\
    PSNR
    \begin{center}
    \textcolor{black}{(f) GT}
    \end{center}
\end{subfigure}%
    \caption{\textbf{Denoising results on the FMD dataset.} Note that for some cases, Self2Self failed to denoise the input image and returned only a blank image.}
\label{fig:FMD}
\end{figure*} 
\textbf{Qualitative and Quantitative Results.} From our experiments we are able to outperform existing state-of-the-art approaches for single image denoising by a significant margin in terms of the PSNR and SSIM. In Table \ref{table:experiment_results}, we can see that for all noise levels in the PINCAT dataset we are significantly outperforming other self-supervised and classical methods in terms of the PSNR and the SSIM. In Figure. \ref{fig:PINCAT}, we show that our approach can recover the finer details around the rib-cage which other methods fail to recover to any meaningful degree. Additionally, we can see that in regions where the ground truth is of a single consistent value we are able to recover the region with a result that is much smoother than other methods where significant noise is still present in the final output. With the FMD dataset where the noise is not pure Poisson noise, we find that on average we are doing better than existing methods in terms of the PSNR. In terms of the SSIM, we are slightly underperforming when compared to the DIP but visual inspection shows that DIP in certain cases DIP will actually over-smooth the resulting image as shown in Figure \ref{fig:FMD}. We additionally found that Self2Self is unstable on many of the images resulting in blank images for a large portion of the dataset which explains the poor results shown in Table \ref{table:experiment_results}. Overall our method performs better than other self-supervised single image approaches for pure Poisson noise and is competitive in cases where the noise is not completely Poisson noise. \\[1em]
\begin{table}[!t]
\begin{minipage}{0.63\textwidth}
\centering
\renewcommand{\arraystretch}{1.2}
\setlength{\tabcolsep}{2pt}
\resizebox{0.96\textwidth}{!}{
\begin{tabular}{c|c|c|c|c|c}
\hline
\multicolumn{2}{c|}{\textbf{Loss Functions}} & \multicolumn{4}{c}{\textbf{Metrics}}
\\ \hline
$\mathcal{L}_{L1}$ & $\mathcal{L}_{Poisson}$ & \textbf{PSNR (dB)} & $\rm{\Delta}$\textbf{PSNR} & ~\textbf{SSIM}~ & $\rm{\Delta}$\textbf{SSIM} \\ \hline
 \checkmark & \checkmark & 32.202 & -- & 0.9370 & -- \\
 \checkmark &  & 31.565 & -2.01 \% & 0.9335 & -0.37 \% \\
    & \checkmark & 32.007 & -0.61 \% & 0.9356 & -0.14 \% \\ 
\hline
\end{tabular}}
\end{minipage}
\begin{minipage}{0.35\textwidth}
\caption{\textbf{Ablation studies.} We tested the impact of each term in our loss function on the overall performance, indicated by a \checkmark.}
\label{table:ablation}
\end{minipage}
\end{table}

\textbf{Ablation Studies.} We performed ablation studies on our final loss function \eqref{eq:full_loss} in order to evaluate the performance of its components. We evaluate our approach using the same architecture for the full and ablated loss functions. Specifically, we test our full loss function, $\mathcal{L}_{Poisson}$, and $\mathcal{L}_{L1}$. For the experiments, we keep the regularizer $\mathcal{L}_N$ for all experiments and tested our results on the PINCAT dataset with Poisson noise with strength $\lambda=20$. We use the same metrics used to evaluate the performance of our approach - PSNR, and SSIM. The results, as shown in Table \ref{table:ablation}, indicate that our Poisson-based loss is the key contributor to the performance of our overall approach; when it is removed we suffer the largest performance drop as opposed to removing the L1 reconstruction loss. This validates our motivation of using a specialized loss function for handling Poisson noise. In addition to this, we find that the inclusion of the L1 reconstruction loss does improve our performance to a level greater than any of the individual terms on their own. Qualitative results of the ablation studies are shown in the supplementary material.

\section{Conclusion}
In this paper, we introduce Poisson2Sparse, a self-supervised approach for single-image denoising for Poisson corrupted images. We explore the application of sparsity, in conjunction with internal learning-based methods for image enhancement, and show significantly superior performance to existing approaches. By only requiring a single noisy image, our method is practical in situations where the acquisition of clean data can be difficult. Our experiments validate our method as we are able to outperform existing state-of-the-art methods for self-supervised approaches under a variety of datasets and \calvin{varying levels of Poisson noise}, for example, by more $\sim 2$ dB PSNR in average performance.


    
\paragraph{Acknowledgement.} The work was partially supported by US National Science Foundation grants 1664172, 1762063, and 2029814.

\FloatBarrier
%
%
%
\bibliographystyle{splncs04}
\bibliography{ms}

\end{document}


\title{Supplementary Material for ``Poisson2Sparse: Self-Supervised Poisson Denoising From a Single Image"}
\author{Calvin-Khang Ta$^{1,}$\thanks{joint first authors},
Abhishek Aich$^{1,\star}$,
Akash Gupta$^{2,\star}$,
Amit K. Roy-Chowdhury$^1$}
\authorrunning{C.K Ta et al.}
\titlerunning{Poisson2Sparse: Self-Supervised Poisson Denoising From a Single Image}
%
\institute{$^1$University of California, Riverside, $^2$Vimaan AI\\
\email{\{cta003@, aaich001@, agupt013@, amitrc@ece.\}ucr.edu}
}
\maketitle              
\begin{algorithm}[h!]
\small{
\caption{Proposed Poisson2Sparse Training}
\label{alg:cap}
\SetKwComment{Comment}{/* }{ */}
\SetKwInput{KwInput}{Input\quad}                
\SetKwInput{KwOutput}{Output~}              
\KwInput {$\noisyxmat$ a noisy image, \textit{N} Number of iterations}
\KwInput{$\bm{\theta}$ encoder-decoder parameters}
\KwOutput{denoised image $\bm{\widehat{X}}$}
\Comment{Sparse Representation Optimization via Forward Pass}

$\bm{\mathrm{A}}_0 \gets Encoder (\noisyxmat)$ \\
\For{$i$ \text{in} N \text{Steps}}
{

    $\bm{X_i} \gets Decoder(\bm{\mathrm{A}}_{i-1})$ \\
    $\bm{\mathrm{A}}_i \gets S(\bm{\mathrm{A}}_{i-1} + Encoder(\noisyxmat- \bm{X_{i}}))$ \\

}
$\bm{\widehat{X}} \gets Decoder(\bm{\mathrm{A}}_N)$\\

\Comment{Dictionary Learning via Back-propagation}

Compute $\mathcal{L}$ and update $\theta$ 
}
\end{algorithm}


\begin{figure*}[h!]
\begin{subfigure}[b]{0.16\textwidth} 
    \centering
    \includegraphics[width=\linewidth,trim={1cm 1cm 1cm 1cm},clip]{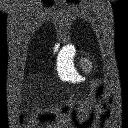}
    16.920 dB
    \includegraphics[width=\linewidth,trim={1cm 1cm 1cm 1cm},clip]{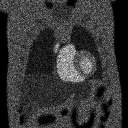}
    17.580 dB
    \begin{center}
    \textcolor{black}{(a) Noisy}
    \end{center}
\end{subfigure}%
\hfill
\begin{subfigure}[b]{0.16\textwidth} 
    \centering
    \includegraphics[width=\linewidth,trim={1cm 1cm 1cm 1cm},clip]{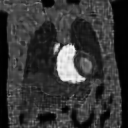}
    21.985 dB
     \includegraphics[width=\linewidth,trim={1cm 1cm 1cm 1cm},clip]{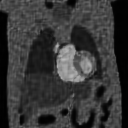}
    23.386 dB
    \begin{center}
    \textcolor{black}{(b) DIP}
    \end{center}
\end{subfigure}%
\hfill
\begin{subfigure}[b]{0.16\textwidth} 
    \centering
    \includegraphics[width=\linewidth,trim={1cm 1cm 1cm 1cm},clip]{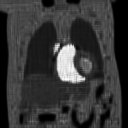}
    26.827 dB
    \includegraphics[width=\linewidth,trim={1cm 1cm 1cm 1cm},clip]{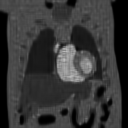}
    27.700 dB
    \begin{center}
    \textcolor{black}{(c) Self2Self}
    \end{center}
\end{subfigure}%
\hfill
\begin{subfigure}[b]{0.16\textwidth} 
    \centering
     \includegraphics[width=\linewidth,trim={1cm 1cm 1cm 1cm},clip]{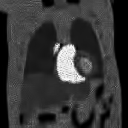}
    28.831 dB
     \includegraphics[width=\linewidth,trim={1cm 1cm 1cm 1cm},clip]{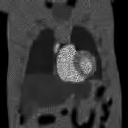}
    29.082 dB
    \begin{center}
    \textcolor{black}{(d) BM3D}
    \end{center}
\end{subfigure}%
\hfill
\begin{subfigure}[b]{0.16\textwidth} 
    \centering
     \includegraphics[width=\linewidth,trim={1cm 1cm 1cm 1cm},clip]{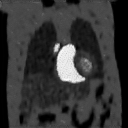}
    \bf{30.433 dB}
        \includegraphics[width=\linewidth,trim={1cm 1cm 1cm 1cm},clip]{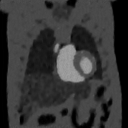}
    \bf{31.494 dB}
    \begin{center}
    \textcolor{black}{(e) Ours}
    \end{center}
\end{subfigure}%
\hfill
\begin{subfigure}[b]{0.16\textwidth} 
    \centering
    \includegraphics[width=\linewidth,trim={1cm 1cm 1cm 1cm},clip]{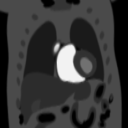}
    PSNR
     \includegraphics[width=\linewidth,trim={1cm 1cm 1cm 1cm},clip]{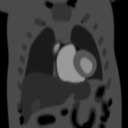}
    PSNR
    \begin{center}
    \textcolor{black}{(f) GT}
    \end{center}
\end{subfigure}%
    \caption{\textbf{Denoising results on the PINCAT dataset}. Noise is $\lambda=20$ on the first row and $\lambda=40$ on the second row.}
\label{fig:PINCAT}
\end{figure*}

\begin{figure*}[h!]
\begin{subfigure}[b]{0.26\textwidth} 
    \centering
    19.346 dB 
    \includegraphics[width=\linewidth]{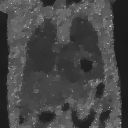}

\end{subfigure}%
\hfill
\begin{subfigure}[b]{0.26\textwidth} 
    \centering
    19.154 dB
    \includegraphics[width=\linewidth]{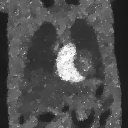}

\end{subfigure}%
\hfill
\begin{subfigure}[b]{0.26\textwidth} 
    \centering
    19.179 dB
    \includegraphics[width=\linewidth]{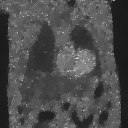}

\end{subfigure}%
\hfill
    \caption{\textbf{Denoising results on the PINCAT dataset}. The following images were denoised using ISTA. Noise is $\lambda=10$.}
\label{fig:PINCAT}
\end{figure*} 

\begin{figure*}[!t]
\begin{subfigure}[b]{0.28\textwidth} 
    \centering
    \caption{Noisy} 
    \includegraphics[width=.9\linewidth,trim={0cm 8cm 8cm 0cm},clip]{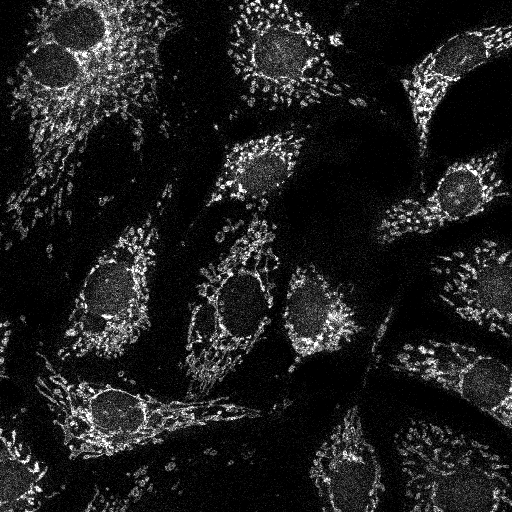}

    \includegraphics[width=.9\linewidth,trim={0cm 8cm 8cm 0cm},clip]{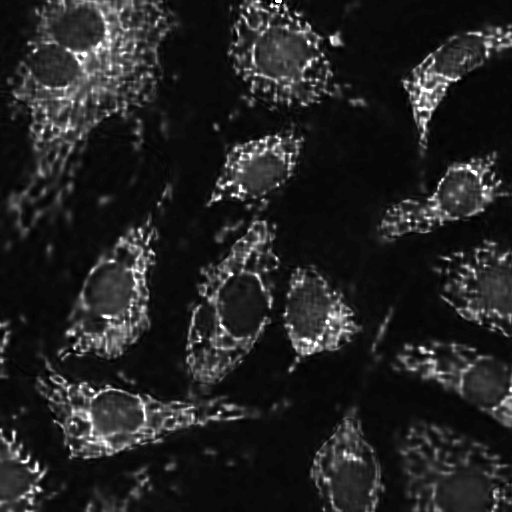}
    \caption{DIP}
\end{subfigure}%
\hfill
\begin{subfigure}[b]{0.28\textwidth} 
    \centering
    \caption{Self2Self} 
    \includegraphics[width=.9\linewidth,trim={0cm 8cm 8cm 0cm},clip]{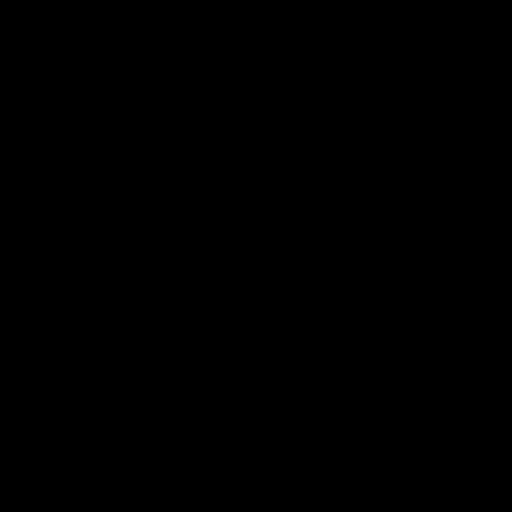}

    \includegraphics[width=.9\linewidth,trim={0cm 8cm 8cm 0cm},clip]{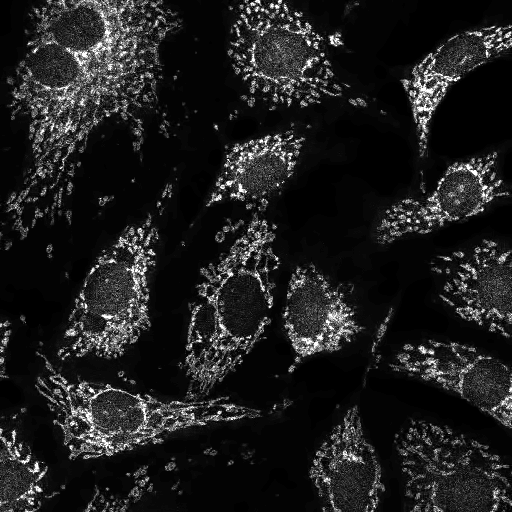}
    \caption{BM3D} 
\end{subfigure}%
\hfill
\begin{subfigure}[b]{0.28\textwidth} 
    \centering
    \caption{\textbf{Ours}} 
    \includegraphics[width=.9\linewidth,trim={0cm 8cm 8cm 0cm},clip]{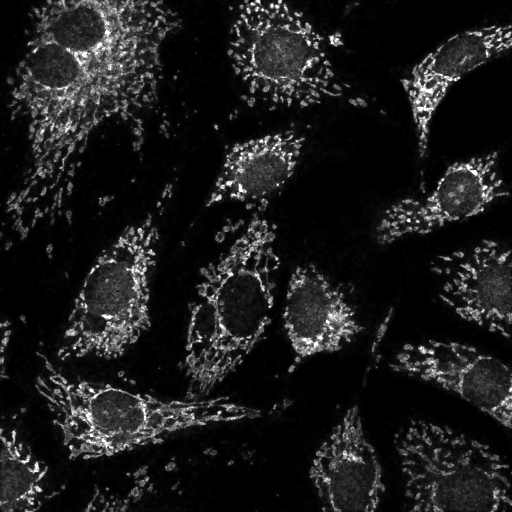}
    \includegraphics[width=.9\linewidth,trim={0cm 8cm 8cm 0cm},clip]{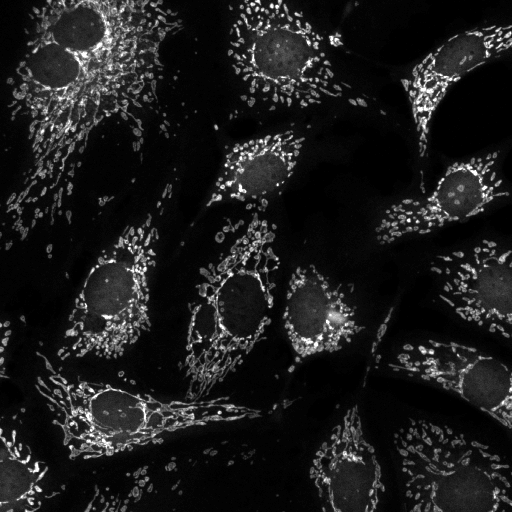}
    \caption{GT} 
\end{subfigure}%
\hfill
    \caption{\textbf{Denoising results on the FMD dataset}. We can see that our result produces the most visually similar image. The DIP will overly smooth the image and you can see that some textural information is loss. Our superior performance is much more apparent when we look at Self2Self and BM3D}
\label{fig:PINCAT}
\end{figure*}